# Polyaniline and graphene nanocomposites for enhancing the interlaminar fracture toughness and thermo-mechanical properties of carbon fiber/epoxy composites

Pralhad Lamichhane[a], Dilli Dhakal[a], Ishan N Jayalath[b], Lynsey Baxter[a], Siddhesh Chaudhari[a], Ranji Vaidyanathan[a]

[a]School of Materials Science and Engineering, Oklahoma State University, Tulsa, OK 74106, USA

[b]Department of Chemistry, Oklahoma State University, Stillwater, OK 74078, USA

**Abstract**

Delamination is one of the major concerns for carbon fiber reinforced polymer (CFRP) composites. Improving fracture toughness reduces delamination and allows for the creation of multifunctional CFRP. In this study, polyaniline (PANI) coated graphene nanoplatelets (GNP) are introduced in the interlaminar region of CFRP pre-preg with the aid of environmentally friendly thermoplastic polymer, polyvinylpyrrolidone (PVP). The GNP content is varied as 1, 3, and 5% by weight with respect to PVP. The improvements in the fracture toughness are investigated using a double cantilever beam test. An improvement of approximately 79% in mode-I fracture toughness is observed with the addition of 5 wt% GNP with respect to PVP. Dynamic mechanical analysis results confirmed the improvements in storage modulus but showed a decrease in glass transition temperature with the addition of nanofiller. Various spectroscopic analysis and imaging techniques were used to understand and evaluate the interactions of nanocomposites with CFRP pre-preg.

## 1. Introduction

Carbon fiber reinforced polymer (CFRP) composites are gaining increasing attention as promising structural materials in the aerospace, automotive, and transportation industries due to their high strength-to-weight ratio. CFRP composites are made of two main components: matrix and reinforcement, where matrix represents polymer (epoxy resin) and the reinforcement represents carbon fiber. The brittle nature of epoxy is a serious problem in CFRP, which results in poor interlaminar fracture toughness and poor out-of-plane performance that ultimately limits the application of CFRP as a multifunctional material. Fracture toughness is an essential parameter in composites that estimates the material's resistance to fracture and has a direct effect on the overall structural performance of fiber-reinforced polymer composites [1]. As a result of crack propagation, delamination of the composite takes place. Therefore, it is crucial to improve fracture toughness and minimize the delamination in CFRP composites to increase their performance and reliability.

Recently, numerous reinforcing techniques have been developed to improve the fracture toughness of laminated composites. The fiber-surface modification, interleaving with a thin film, and three-dimensional interlaminar reinforcement (stitching, z-pinning, weaving) are widely adopted methods [2]. Conversely, these methods may affect the performance of in-plane yarn fibers resulting in degradation of in-plane mechanical properties [3]. Another approach includes matrix modification by the addition of micro-/ nano-fillers, which improve the interlaminar bonding between the matrix and fiber [4, 5]. It has been reported that the addition of nano-fillers such as carbon nanotubes (CNTs), graphene and its derivatives, fullerenes, and nanoclay can significantly increase the strength and fracture toughness of fiber-reinforced composites [6-9]. The use of micro-fillers counteracts crack propagation by crack deflection, pinning, and bowing, while nano-fillers control the fracture behavior by introducing additional modes of energy dissipation during fracture. Graphene and CNTs are promising nano-fillers for matrix modification of CFRP composites.

Graphene and CNTs are superior to other nano-fillers due to their low densities and greater potential to improve mechanical, thermal, and electrical properties [10]. Graphene shows a higher degree of mechanical bridging, such as interlocking at the interface due to its enhanced specific surface area [11] and significant improvements in fracture toughness even in low loadings (e.g., 0.1 wt%) [12]. The toughening mechanisms of CNTs include CNT pull-out and CNT bridging, resulting from an interfacial bond between the filler and resin [13, 14]. Furthermore, promising enhancements to fracture toughness have been reported using graphene oxide between pre-preg layers [15, 16]. Despite these advantages, the issue of non-uniform dispersion and aggregation of nanofiller in the resin system is a considerable concern. Due to van der Waal's forces and their high surface areas, graphene and CNTs tend to agglomerate within the resin system. Such agglomerates create local stresses in response to external loads on the composite, ultimately leading to the agglomerates acting as points of failure in the structure [17]. The modification of the graphene and CNTs with surface functional groups is an effective approach that increases resin compatibility. Surface modifications help to prevent the re-stacking of graphene from van der Waal's forces.

Polyaniline (PANI) is a conducting polymer that has been tested as a successful nanofiller for epoxy matrix modification. Some work has shown that interaction between PANI and epoxy enhances the electrical conductivity and mechanical properties of the carbon fiber and epoxy composites [18-20]. It has been studied that the PANI molecules sandwich between basal planes of graphene nanoplatelets (GNP) due to strong π-π interactions during in-situ polymerization. As a result of PANI molecules, the graphene sheet exfoliates and reduces van der Waal's attractions and ultimately facilitates the dispersion of GNP in polymer matrices [21]. Duan et al. studied the effect of uniform PANI coating on the GNP surface for reducing agglomeration in resin [22]. Studies have reported that PANI could serve as a coupling agent to improve the dispersion quality of nanofillers within polymer matrices and enhance interfacial adhesion by forming covalent bonds between PANI and polymer resin [23, 24].

Polyvinylpyrrolidone (PVP) is an amphiphilic water-soluble polymer that exhibits excellent solubility, good chemical stability, film-forming properties, and increased bonding capacity in composites. The use of PVP as a compatibilizer has shown promising results in increasing fiber-matrix bonding, which improves resultant mechanical properties [25, 26]. Recent studies have showcased some of the appealing characteristics of PVP, such as its outstanding cross-linking capacity and compatibility in most polymer resins. A study by Oyama et al. demonstrated the strong chemical interactions between the epoxy and carboxyl functional groups of PVP [27]. Additionally, introducing PVP into PANI-GNP enhances the dispersion of filler by adsorbing on their surfaces and improving adhesion [28, 29].

This study presents the development of PANI coated GNP nanoparticles by in-situ polymerizations of aniline in the presence of graphene nanoplatelets. The additive was chemically characterized by Fourier-transform infrared spectroscopy (FTIR), scanning electron microscopy (SEM), and x-ray diffraction (XRD) analysis. The PANI coated GNP was mixed with PVP, and the resultant blend was applied between carbon fiber pre-preg layers to analyze the effect on fracture toughness. PANI was selected as a GNP modifier due to its large-scale availability, low cost, and compatibility with PVP and epoxy. The mode-I interlaminar fracture toughness of the modified composites was studied with double cantilever beam (DCB) testing, and the fracture surfaces were analyzed using scanning electron microscopy (SEM). The thermo-mechanical properties of the composites were studied by dynamic mechanical analysis (DMA) testing.

## 2. Experimental procedure
### 2.1 Materials

Graphene nanoplatelets (average surface area 120 - 150 $m^2$/ g, and average particle diameter of 5 µm (Grade M particles)) were purchased from XG Science. PVP (average molecular weight 360,000), sulfuric acid ($H_2SO_4$), hydrochloric acid (HCl), aniline (99.5 % pure), and ammonium persulfate (APS) were purchased from Sigma-Aldrich. Carbon fiber pre-preg 3K, 2 × 2 weave with epoxy resin was bought from Fibre Glast. All the fiber materials and chemical reagents were used without further modification or purification.

### 2.2 Synthesis of PANI-GNP nanocomposites

The PANI-GNP nanocomposites were synthesized by in-situ anionic polymerization of aniline as described in previously reported work [30]. In a typical procedure, 0.47 g GNP was dispersed in 100 mL of 1M HCl using ultrasonication and magnetic stirring. After 20 min, 1.92 g of aniline monomer was added into GNP dispersion with continuous stirring. The pre-cooled APS solution (4.56 g APS dissolved in 100 mL of 1M HCl) was added dropwise into the aforementioned mixture. The reaction was carried out under constant stirring for 4 h in an ice bath. After 4 h, the resulting dark green mixture was filtered and washed with deionized water, HCl, and acetone several times to remove unreacted aniline. Finally, the product was dried in an oven at 80 °C for 10 h to obtain a PANI coated GNP nanocomposite. The weight ratio of GNP: PANI = 1: 4 was measured in the final product of PANI-GNP nanocomposites. The probable reaction mechanism of cationic polymerization of aniline in the presence of HCl is shown in Fig. 1.

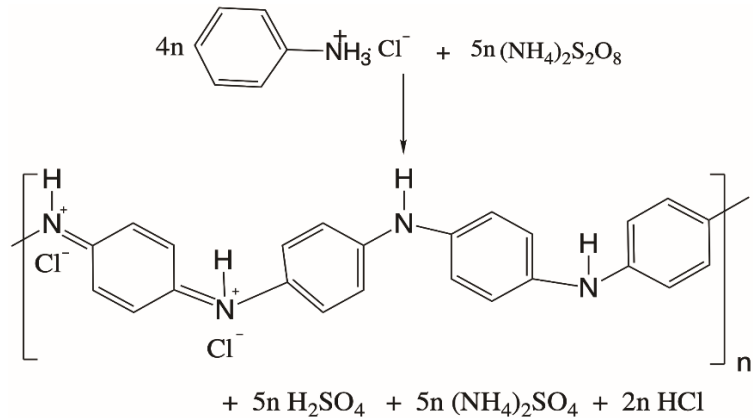

**Fig. 1.** Oxidative polymerization of aniline in the presence of APS

*2.3 Preparation of PVP and PANI-GNP blends*

The PVP and PANI-GNP blend was prepared in an ethanol solution. 5 wt% of PVP was dissolved in ethanol, and a calculated amount of PANI-GNP nanocomposite was added to the PVP solution so that the GNP remained as 1, 3, and 5 wt% with respect to PVP. The composite blend is referred to as PA/PVP-GNP throughout this literature. To prepare the PA/PVP sample, pure PANI (10 wt% with respect to PVP) was dissolved in the PVP solution to prepare a PVP and PANI blend.

*2.4 Coating and Fabrication of composites*

The interlaminar surface of CFRP pre-pregs was modified using the PVP and PANI-GNP blend. Two perpendicular layers of the homogeneous PA/PVP-GNP blend were uniformly applied on the mid-plane of the 16 layered CFRP laminates with the help of a paintbrush. A Teflon release film was placed across the width of the mid-plane to produce an initial delamination zone for the DCB tests. The laminated composites were then placed between two Teflon sheets and cured in a hot press. Constant pressure was applied uniformly and heated throughout the curing process according to the supplier's specifications. The composite panels were heated using a 2 °C/ min ramp up to 154 °C, then maintained for 1 h before cooled down to 66 °C at 2 °C/ min. The neat CFRP pre-preg was considered as a control sample.

Separate test panels were made for the dynamic mechanical analysis (DMA). For each variation of GNP, PA/PVP-GNP blend was applied between every layer of CFRP pre-preg. A similar procedure as above was used for curing the DMA samples.

*2.5 Characterization*
*2.5.1 Morphology and fracture surface analysis*

The surface morphology of the GNP, PANI, PANI-GNP and the fracture surface of modified CFRP laminates were investigated using an FEI Quanta 600 field-emission gun environmental scanning electron microscope (FESEM) (Waltham, MA). The images of the fractured surfaces provide further insight into potential toughening mechanisms promoted by these modifications.

*2.5.2 FTIR*

Fourier-transform infrared spectroscopy was used to investigate the chemical interaction between PANI and GNP. GNP, PANI, and PANI-GNP nanocomposite powders were added to KBr in a ratio of 1:85. These powdered samples were loaded in a Nicolet iS10 FTIR spectrometer (Waltham, MA). The FTIR spectra of composites were collected with 64 scans and 4.0 cm$^{-1}$.

### 2.5.3 XRD

XRD spectra were collected using a Bruker AXS D8 Discover X-ray Diffractometer with a general area detector diffraction system (GADDS) Vantec 500 2D detector. The wavelength used was 1.54056 Å, and with a scanning angle of 2θ from 10° to 40°.

### 2.5.4 Mode-I interlaminar fracture toughens

The interlaminar fracture toughness of the composites was tested using DCB testing according to ASTM D-5528. All the test samples were cut and prepared to meet the ASTM standard with 135 mm in length and 20-25 mm in width. The test set up for the DCB test using a Universal Testing Machine (Instron-5582) is shown in Fig. 2. The energy release rate under mode I conditions ($G_I$) was calculated using the modified beam theory and shown in Equation 1.

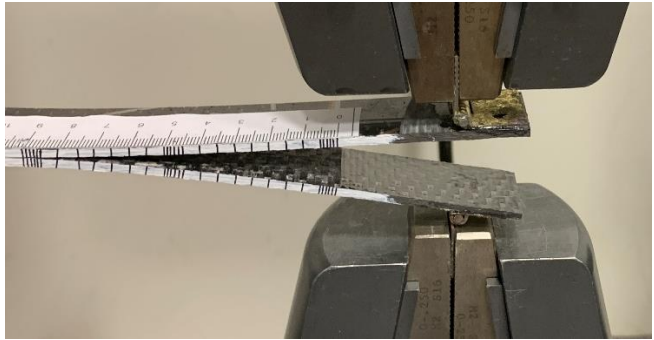

**Fig. 2.** Test setup for DCB test

$$G_{IC} = \frac{3P\delta}{2b(a+|\Delta|)} \quad (1)$$

where,
P = load, δ = load point displacement, b = specimen width, a = delamination length, |Δ| = correction factor. The correction factor is calculated by plotting $C^{1/3}$ vs. a, where C is the ratio of crack mouth opening displacement (CMOD) to delamination length (a). The load-displacement data and crack propagation length was recorded and used for the calculation of $G_{IC}$.

### 2.5.5 Dynamic Mechanical Analysis (DMA)

The thermomechanical properties of the composite laminates were studied using DMA tests (TA Instruments, DMA Q800). The three-point bending test was conducted while simultaneously heating the sample from room temperature to 125 °C. The ramp rate was maintained at 3 °C/min applying a constant sinusoidal displacement of 20 μm with a 1 Hz frequency. The storage modulus, loss modulus, and damping factor were obtained across the temperature range.

## 3. Results and Discussion
*3.1. FTIR*

The chemical characterization of GNP, PANI, and PANI-GNP was carried out using FTIR analysis, and spectra are shown in Fig. 3a. FTIR spectrum of pure GNP exhibited bands at 3445 cm$^{-1}$, 1635 cm$^{-1}$, 1265 cm$^{-1}$, 1097 cm$^{-1}$, and 802 cm$^{-1}$, indicating the stretching vibrations of O-H, C=C stretching vibrations, C-OH stretching vibrations, C-O stretching vibrations, and C-H bends vibrations, respectively [31, 32]. For pure PANI, the broadband at 3425 cm$^{-1}$ is attributed to N-H stretching vibrations [33], and bands located at 1563 cm$^{-1}$ and 1466 cm$^{-1}$ are ascribed to C=C stretching vibrations for the quinoid ring and benzenoid ring, respectively [32]. The peaks at 2919 cm$^{-1}$, 1302 cm$^{-1}$, 1115 cm$^{-1}$, and 796 cm$^{-1}$ represent C-H bending vibrations, aromatic amine C-N stretching vibrations, and in-plane aromatic C-H bending vibrations, and out-of-plane aromatic C-H bending vibrations, respectively [32, 34].

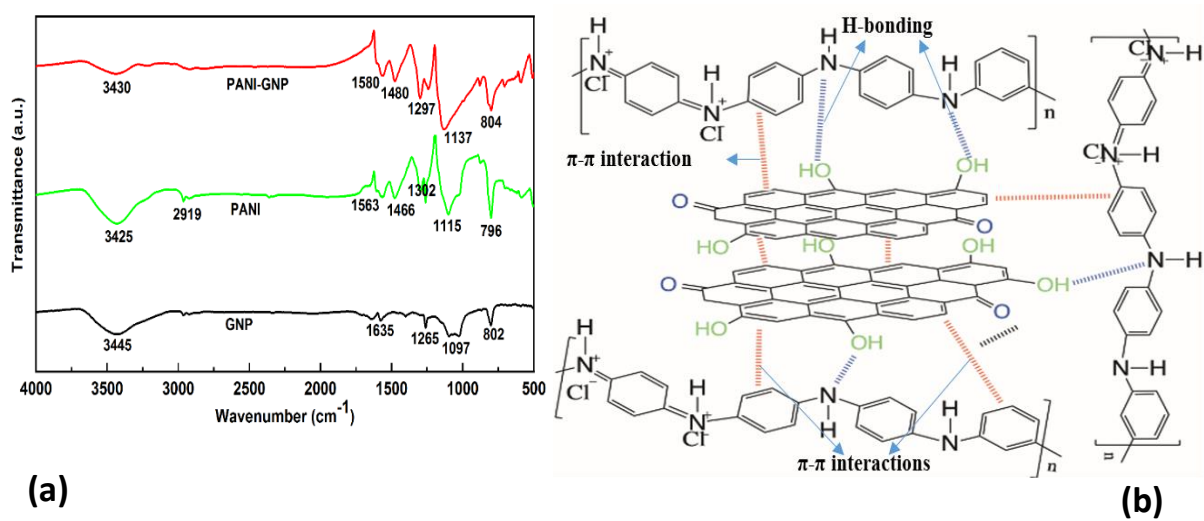

**Fig. 3.** (a) FTIR spectra of PANI, GNP, and PANI coated GNP (b) the schematic representation of possible interactions between graphene sheets and PANI molecules

FTIR can be used to study the interaction between PANI and graphene sheets. The presence of -OH, -C=C, -C-OH, and -C-O groups on the GNP surface directly interacts with PANI by electrostatic attraction and the formation of hydrogen bonds [35, 36]. The interactions between the PANI and GNP are shown in Fig. 3b. PANI coated GNP nanocomposites showed similar absorption peaks as PANI with slightly shifted positions (as seen in Fig. 3a 3440, 1580, 1480, 1297, 1137, and 804 cm$^{-1}$). The broadening and increased intensities were also observed in PANI-GNP nanocomposites, showing the PANI and GNP interactions. The band's shift of stretching vibrations of C=C to a higher wavenumber demonstrates the π-π interaction existence and formation of hydrogen bonds between GNP and PANI [37].

*3.2. XRD*

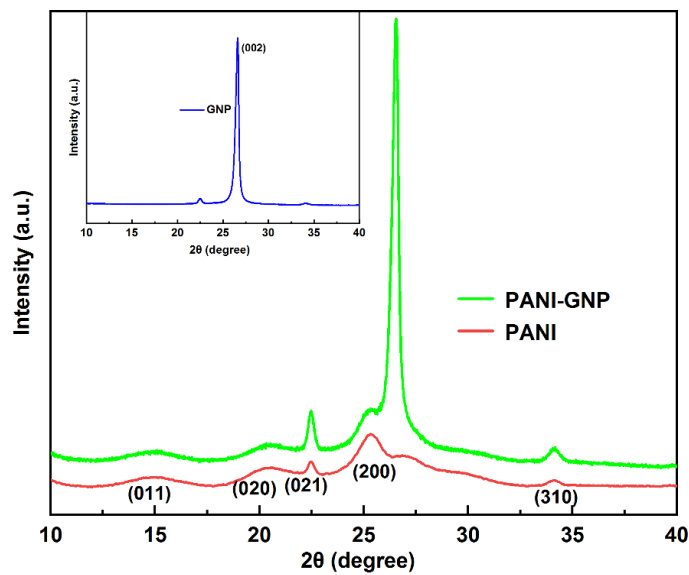

**Fig. 4.** XRD patterns of PANI and PANI coated GNP; inset is XRD of pure GNP

The XRD spectra of GNP, PANI, and PANI- GNP nanocomposites are shown in Fig. 4. PANI exhibited the major characteristic peaks at 15.03°, 20.31°, 22.46°, 25.34°, and 34.15° of 2θ corresponding to diffraction planes (011), (020), (021), (200), and (310), respectively that are comparable with previous work [38, 39]. PANI is a 2-phase system as it contains benzenoid and quinoid groups. The phase in which the polymer chains are perfectly ordered in the close-packed array is the crystalline region, while the chains are randomized in the amorphous area [40]. Based on the diffraction patterns, it can be seen that the synthesized polyaniline has a more ordered structure with semi-crystalline nature. GNP has a single high intense characteristic peak observed at 2θ of 26.5°, corresponding to the (002) crystal plane of graphite nanoplatelets [41]. The PANI-GNP nanocomposites demonstrate diffraction peaks corresponding to both PANI and GNP. This result shows the presence of both PANI and GNP in the composite and successful in-situ synthesis of PANI on the GNP surface.

*3.3. SEM*

The scanning electron microscopy images of pure GNP, PANI, and PANI coated GNP is shown in Fig. 5. GNP (Fig. 5a.) was observed as a flaky, translucent, and relatively smooth surface having small folds with no other substances attached to the surface. The morphology of PANI (Fig. 5b) has appeared as granular round-shaped structures. The micro-granular PANI structure has formed by the aggregation that results from interchain interactions of small globular structures. The rough surface (Fig. 5c) of PANI-GNP was observed compared to that of GNP, indicating the presence of PANI on the GNP surface. The small, thickly accumulated, rough substances on the GNP show that PANI is coated on the GNP surface during the polymerization.

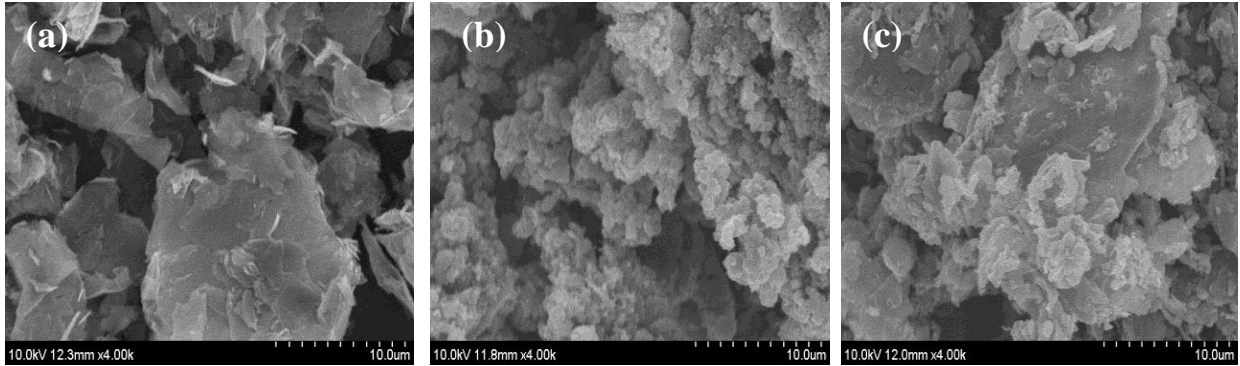

**Fig. 5.** SEM images of (a) GNP, (b) PANI, and (c) PANI coated GNP

*3.4. DCB*

*3.4.1. Fracture toughness measurements*

Mode-I interlaminar fracture toughness of composites was studied by load-displacement graphs, and the results are shown in Fig. 6. The corresponding R-curves for all the samples are presented in Fig. 7c. The load has increased linearly in the linear elastic region in each sample until the crack initiation point. Then, the load decreased gradually once the crack propagated further in the composite. The curves are not continuous; however, they are a sequence of arrests that manifest the characteristics of "stick-slip" behavior every time the crack propagates. The maximum load was increased for each nanofiller added composites, indicating greater energy required for crack propagation. This is attributed to the nanofiller-fiber bridging and the formation of mechanical bonds between the matrix and fiber. This can be explained by the observation of fiber pull-out during the DCB test for higher GNP-loaded composites (Fig. 7b) compared to the control sample (Fig. 7a).

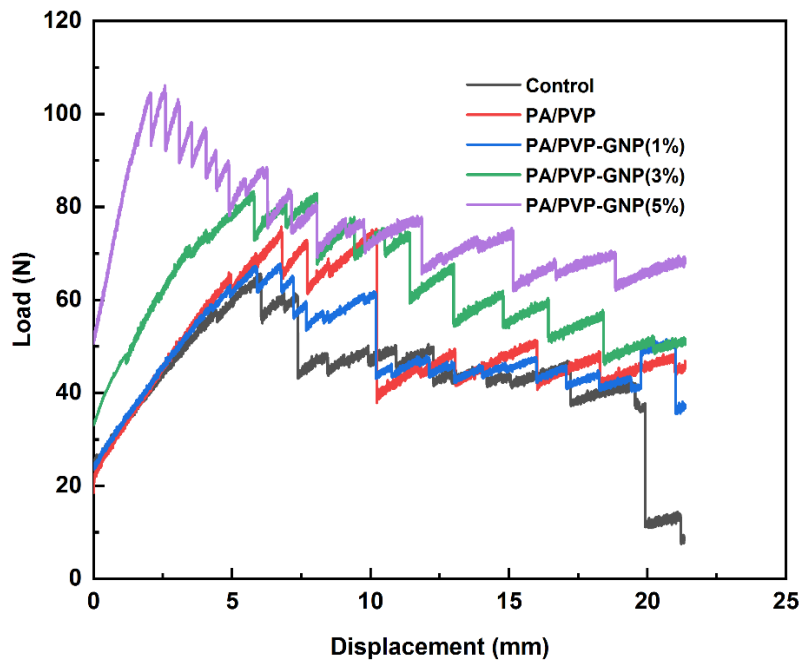

**Fig. 6.** Load-Displacement curves of different composites

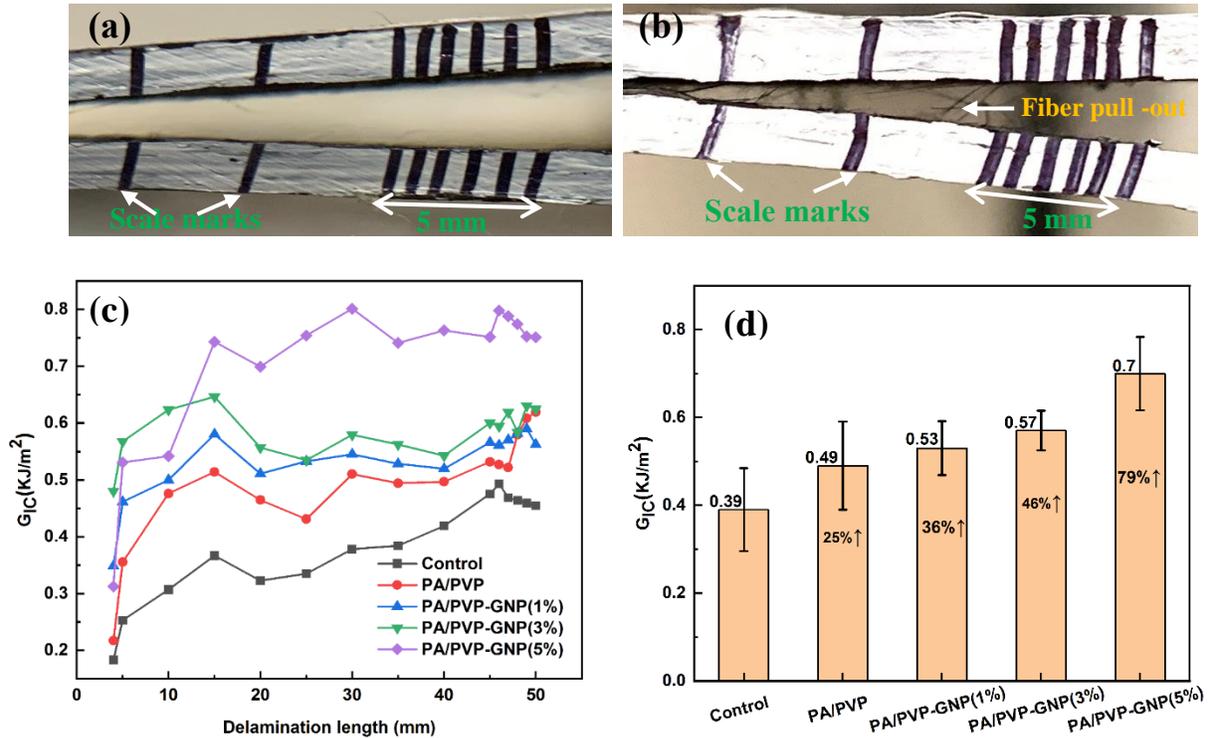

**Fig. 7.** DCB fracture optical images of (a) control sample and (b) PA/PVP-GNP (5%) nanofiller added sample, (c) R-curves for control and nanofiller added composites, (d) Fracture toughness of different composites

Mode-I fracture toughness of control and all nanofiller-loaded composites are shown in Fig.7d. The fracture toughness of nanofiller-added composites was higher compared to the control sample. The $G_{IC}$ for the control sample was 0.39 KJ/m$^2$, whereas $G_{IC}$ in the presence of PANI and PVP nanofiller is 0.49 KJ/m$^2$ approximately a 25% increase. The fracture toughness was increased with the higher loading of GNP in the composites. The $G_{IC}$ for PA/PVP-GNP (5%) composite was measured as 0.7 KJ/m$^2$ approximately a 79% increase compared to the control sample. The increase in fracture toughness can be attributed to the presence of physical attractions such as H-bonding, π- π interactions, and van der Waal's attraction between the epoxy matrix and the carbon fiber due to different functional groups present in GNP and PVP. A similar effect was observed by Mishra et al. using PVP and GO in CFRP composites, as their study found a 33% increase in toughness by the addition of PVP alone [16]. Previous studies have shown that the addition of thermoplastic sizing could modify the physical properties of the interphase region and improve the fatigue and tensile strength of composites [42, 43]. PVP acts as a stabilizer for filler and affects the mechanical properties of composites by establishing a greater degree of contact between the fiber and matrix, and strong fiber/matrix adhesion can be formed. The actual mechanism for the bond formation between the PVP with PANI and GNP is still under study. It could possibly be evaluated using theoretical calculations such as density functional theory and nuclear magnetic resonance.

3.4.2. SEM fractographic

The SEM fractographic of different samples after the DCB tests are shown in Fig. 8. The control sample has a smooth fracture surface showing limited river-like markings. This type of fracture is associated with the brittle nature of epoxy which infers low resistance for crack propagation [44]. In addition, some parts of the cracked surfaces show a gap between fiber and matrix caused by the delamination due to weak interfacial adhesion. On the other hand, all PA/PVP-GNP composites showed relatively rough matrix fractures compared to the control sample which indicates the more ductile failure of the PA/PVP-GNP composites. The PA/PVP composite showed more roughness, crack arresting, and jumping patterns as observed in Fig. 8b. Similarly, all PA/PVP-GNP composites showed many small fracture surfaces along with the crack jumping and stopping configurations. The small fracture surfaces represent the additional failure mechanisms during the crack propagation process due to GNP addition, while the crack jumping and arresting are attributed to the high crack deflection. It is also worth noting that fiber pull-out and fiber breakage, observed in Fig. 8 (c, d, and e), are attributed to the strong adhesion between the fiber and matrix that ultimately lead to high fracture toughness with the addition of nano-fillers.

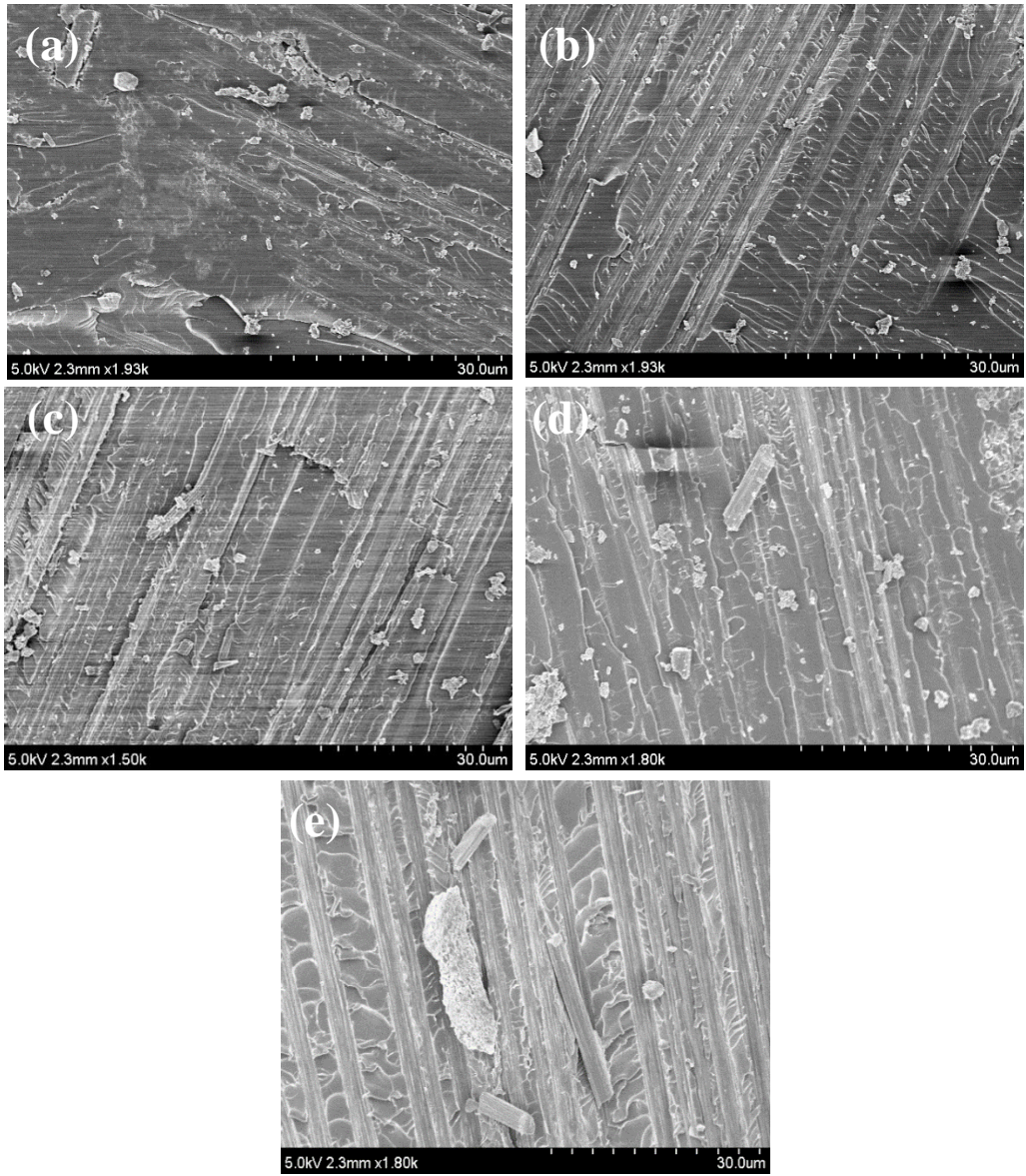

**Fig. 8.** SEM images of fracture surfaces: (a) Control, (b) PA/PVP, (c) PA/PVP-GNP (1%), (d) PA/PVP-GNP (3%) (e) PA/PVP-GNP (5%)

*3.4. Dynamic mechanical analysis*

The thermo-mechanical behavior of the control and all PA/PVP-GNP composites was studied by DMA analysis. Fig. 9 shows the storage modulus and loss modulus of all samples as a function of temperature. In the PA/PVP samples, storage modulus is reduced compared to the control sample, indicating that PVP with PANI introduces plasticization in the composites. However, adding GNP to the composites enhances the storage modulus obtaining a maximum storage modulus at 5 wt% loadings of GNP, with a 26% increase compared to the control sample. This can be due to the mechanical interlocking developed by GNP. In addition, GNP makes epoxy stiffer and shows a higher storage modulus due to the inhibiting of chain mobility in these samples. The room temperature storage modulus represents the stiffness of a viscoelastic material, and many reports claimed stiffness increases within CFRP composites with added graphene [45, 46].

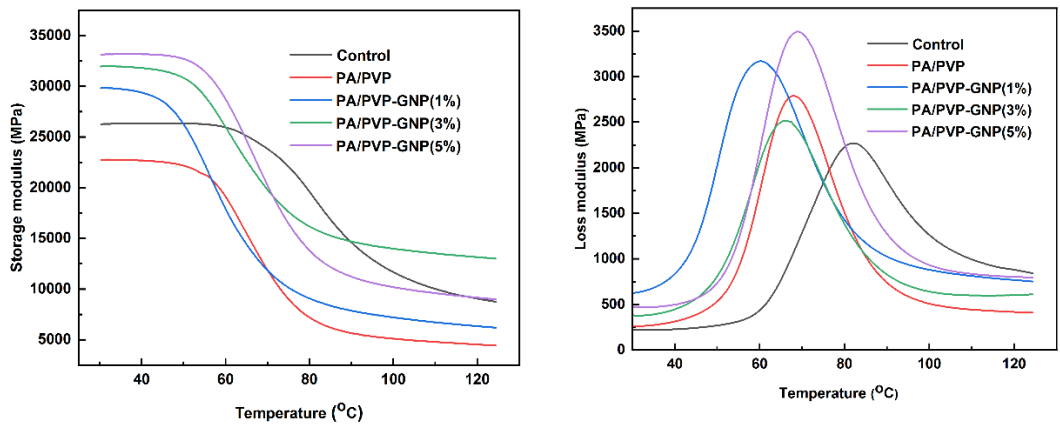

**Fig. 9.** Storage modulus and loss modulus of composites

The storage modulus reaches its maximum value at room temperature and decreases continuously with increasing temperature for all samples. As temperature increases, the mobility of polymer chains increases in the matrix and decreases the effective stress transfer between the fiber and matrix. As a result, the stiffness of viscoelastic materials gradually decreases, showing a downward movement in the storage modulus curve. Interestingly, the control sample showed a higher storage modulus of the slight plateau at a higher temperature region (60 - 100 °C) compared to the other composites. This is because of the inclusion of nanofiller with PVP, which increases the size of the polymer network, supports the polymer chain's motion, and readily decreases the storage modulus when the temperature rises [16, 47].

Table 1. Thermo-mechanical properties of CFRP composites

| Sample types | Storage modulus (MPa) | Change % | $T_g$ (°C) | Change (°C) |
|---|---|---|---|---|
| Control | 26263 ± 1234 | … | 85.94 ± 2.14 | … |
| PA/PVP | 22744 ± 1123 | -13.4 | 66.07 ± 1.45 | -19.87 |
| PA/PVP-GNP (1%) | 29825 ± 1324 | 13.6 | 67.51 ± 3.24 | -18.43 |
| PA/PVP-GNP (3%) | 31963 ± 1231 | 21.7 | 69.3 ± 3.12 | -16.64 |
| PA/PVP-GNP (5%) | 33122 ± 1029 | 26.2 | 74.36 ± 2.45 | -11.58 |

Fig. 10 shows the change in the tan delta value of composites with temperature. Tan delta is known as the damping ratio of loss modulus to storage modulus and provides information regarding the viscous properties of materials. The glass transition temperature ($T_g$) of the composites is the portion of a sharp drop of storage modulus and relates to the peak of the tan delta curve. As shown in Fig. 11, the $T_g$ of the nanofiller added samples decreases, with a minimum temperature of 66.07 °C in the PA/PVP sample compared to the control sample, which also supports the relatively abrupt decline of the storage modulus curve with increasing temperature. This decrease in $T_g$ may be the effect of PVP and PANI that separates the polymer chain, and the chain-to-chain interaction is minimized at a higher temperature. These physical defects may hinder the curing reaction of epoxy and lead to a decrease in crosslinking density, as reported in the previous research literature [48-50]. However, the addition of GNP restricted the polymer chain mobility, and an increase in $T_g$ was observed with increasing the GNP loading, 74.36 °C at 5% GNP.

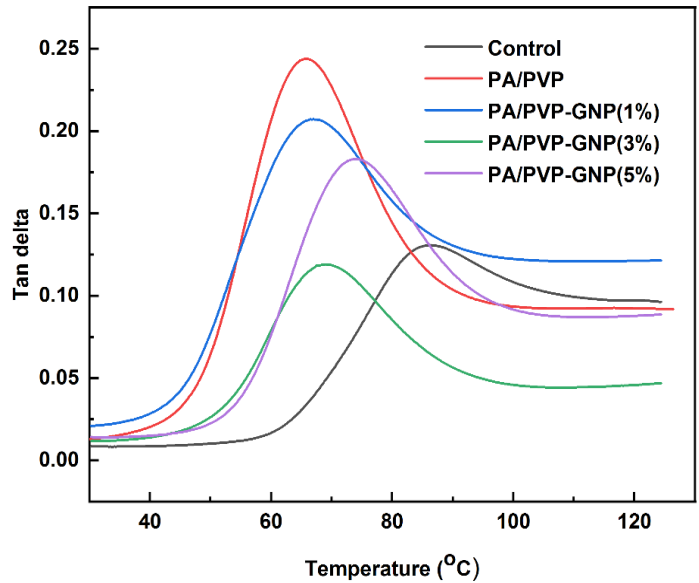

**Fig. 10**. Tan delta with response to temperature

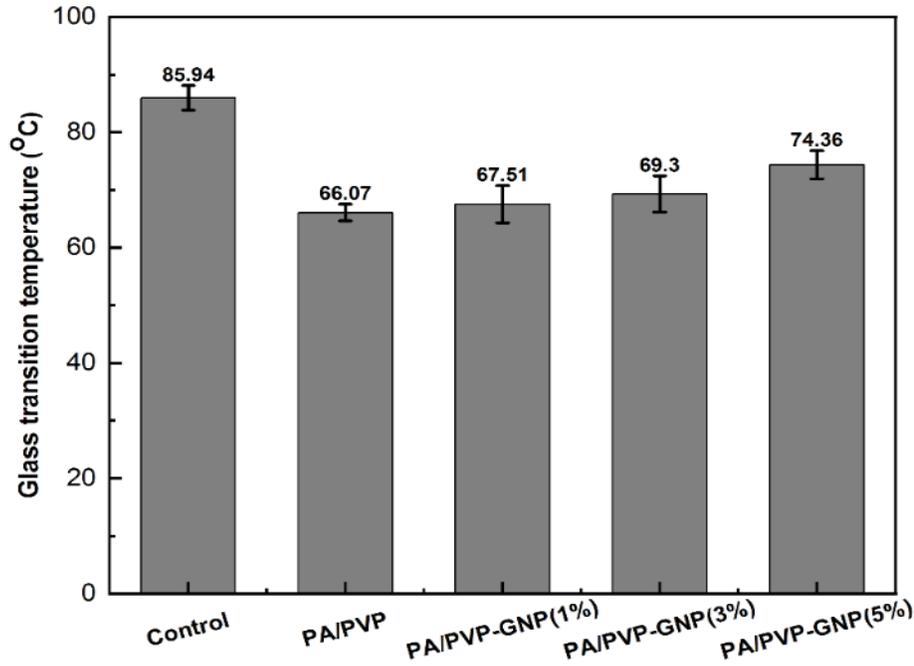

**Fig. 11.** The glass transition temperature of different composites

### 4. Conclusions

Polyaniline coated graphene nanoplatelets were successfully fabricated by in situ polymerizations of aniline. The presence of PANI on GNP surfaces was confirmed by FTIR, XRD, and SEM analysis. The nanocomposites were incorporated on CFRP pre-preg layers with the help of PVP and ethanol. The fracture surface analysis and DCB analysis confirmed the enhancement of mode-I fracture toughness by the addition of nanofiller on CFRP pre-preg. There was a 79% improvement in mode-I fracture toughness in the 5 wt% GNP composite samples. The toughening mechanism was attributed to the formation of covalent bonds between PANI and epoxy, H-bonding between fillers, and crack branching and deflection due to the addition of GNP. DMA tests showed the enhancement of storage modulus by the addition of GNP, while the PVP and PANI filler reduced it due to the development of the thermoplastic property. Overall, the PANI coated GNP aids in the enhancement of fracture toughness as well as other mechanical properties in CFRP pre-preg laminate composites.

**Acknowledgment**

This work was financially supported by the National Science Foundation I-Corps site [project number 1548003]; and Helmerich family endowment funds through the Varnadow Chair funds to Dr. Ranji Vaidyanathan. Lisa Whitworth collected the XRD spectra at the OSU microscopy laboratory, Stillwater.

# References

1. Kim, J.-K. and Y.-W. Mai, *High strength, high fracture toughness fibre composites with interface control—a review.* Composites Science and Technology, 1991. **41**(4): p. 333-378.
2. Nagi, C.S., et al., *Spray deposition of graphene nanoplatelets for modifying interleaves in carbon fibre reinforced polymer laminates.* Materials & Design, 2020. **193**: p. 108831.
3. Mouritz, A., *Review of z-pinned composite laminates.* Composites Part A: applied science and manufacturing, 2007. **38**(12): p. 2383-2397.
4. Gojny, F.H., et al., *Influence of nano-modification on the mechanical and electrical properties of conventional fibre-reinforced composites.* Composites Part A: Applied Science and Manufacturing, 2005. **36**(11): p. 1525-1535.
5. Siddiqui, N.A., et al., *Mode I interlaminar fracture behavior and mechanical properties of CFRPs with nanoclay-filled epoxy matrix.* Composites Part A: Applied science and manufacturing, 2007. **38**(2): p. 449-460.
6. Wichmann, M.H., et al., *Glass-fibre-reinforced composites with enhanced mechanical and electrical properties–benefits and limitations of a nanoparticle modified matrix.* Engineering Fracture Mechanics, 2006. **73**(16): p. 2346-2359.
7. Haque, A., et al., *S2-glass/epoxy polymer nanocomposites: manufacturing, structures, thermal and mechanical properties.* Journal of Composite materials, 2003. **37**(20): p. 1821-1837.
8. Karapappas, P., et al., *Enhanced fracture properties of carbon reinforced composites by the addition of multi-wall carbon nanotubes.* Journal of Composite Materials, 2009. **43**(9): p. 977-985.
9. Sager, R.J., et al., *Interlaminar fracture toughness of woven fabric composite laminates with carbon nanotube/epoxy interleaf films.* Journal of applied polymer science, 2011. **121**(4): p. 2394-2405.
10. Atif, R., I. Shyha, and F. Inam, *Mechanical, thermal, and electrical properties of graphene-epoxy nanocomposites—A review.* Polymers, 2016. **8**(8): p. 281.
11. Ferrari, A.C., et al., *Science and technology roadmap for graphene, related two-dimensional crystals, and hybrid systems.* Nanoscale, 2015. **7**(11): p. 4598-4810.
12. Rafiee, M.A., et al., *Enhanced mechanical properties of nanocomposites at low graphene content.* ACS nano, 2009. **3**(12): p. 3884-3890.
13. Shtein, M., et al., *Fracture behavior of nanotube–polymer composites: Insights on surface roughness and failure mechanism.* Composites science and technology, 2013. **87**: p. 157-163.
14. Gojny, F.H., et al., *Influence of different carbon nanotubes on the mechanical properties of epoxy matrix composites–a comparative study.* Composites science and technology, 2005. **65**(15-16): p. 2300-2313.
15. Ning, H., et al., *Interlaminar mechanical properties of carbon fiber reinforced plastic laminates modified with graphene oxide interleaf.* Carbon, 2015. **91**: p. 224-233.
16. Mishra, K., et al., *Effect of graphene oxide on the interlaminar fracture toughness of carbon fiber/epoxy composites.* Polymer Engineering & Science, 2019. **59**(6): p. 1199-1208.
17. Iqbal, K., et al., *Impact damage resistance of CFRP with nanoclay-filled epoxy matrix.* Composites Science and Technology, 2009. **69**(11-12): p. 1949-1957.
18. Zhang, X., et al., *Flame-retardant electrical conductive nanopolymers based on bisphenol F epoxy resin reinforced with nano polyanilines.* ACS applied materials & interfaces, 2013. **5**(3): p. 898-910.
19. Guo, J., et al., *Significantly enhanced mechanical and electrical properties of epoxy nanocomposites reinforced with low loading of polyaniline nanoparticles.* RSC advances, 2016. **6**(25): p. 21187-21192.

20. Lamichhane, P., et al., *Polyaniline doped graphene thin film to enhance the electrical conductivity in carbon fiber-reinforced composites for lightning strike mitigation.* Journal of Composite Materials, 2021: p. 00219983211041751.
21. Xiang, J. and L.T. Drzal, *Templated growth of polyaniline on exfoliated graphene nanoplatelets (GNP) and its thermoelectric properties.* Polymer, 2012. **53**(19): p. 4202-4210.
22. Duan, Y., et al., *First-principles calculations of graphene-based polyaniline nano-hybrids for insight of electromagnetic properties and electronic structures.* RSC advances, 2016. **6**(77): p. 73915-73923.
23. Gu, H., et al., *Polyaniline stabilized magnetite nanoparticle reinforced epoxy nanocomposites.* ACS applied materials & interfaces, 2012. **4**(10): p. 5613-5624.
24. Zhang, X., et al., *Polyaniline stabilized barium titanate nanoparticles reinforced epoxy nanocomposites with high dielectric permittivity and reduced flammability.* Journal of Materials Chemistry C, 2013. **1**(16): p. 2886-2899.
25. Yoon, S. and I. In, *Role of poly (N-vinyl-2-pyrrolidone) as stabilizer for dispersion of graphene via hydrophobic interaction.* Journal of materials science, 2011. **46**(5): p. 1316-1321.
26. Possart, W., *Adhesion: current research and applications.* 2006.
27. Oyama, H.T., J. Lesko, and J. Wightman, *Interdiffusion at the interface between poly (vinylpyrrolidone) and epoxy.* Journal of Polymer Science Part B: Polymer Physics, 1997. **35**(2): p. 331-346.
28. Wei, W., W. Lu, and Q.-h. Yang, *High-concentration graphene aqueous suspension and a membrane self-assembled at the liquid–air interface.* Carbon, 2011. **49**(8): p. 2879.
29. Wajid, A.S., et al., *Polymer-stabilized graphene dispersions at high concentrations in organic solvents for composite production.* Carbon, 2012. **50**(2): p. 526-534.
30. Dhakal, D.R., et al., *Influence of graphene reinforcement in conductive polymer: Synthesis and characterization.* Polymers for Advanced Technologies, 2019. **30**(9): p. 2172-2182.
31. Song, Y., et al., *Enhancing the thermal, electrical, and mechanical properties of silicone rubber by addition of graphene nanoplatelets.* Materials & Design, 2015. **88**: p. 950-957.
32. Wang, Q., et al., *High‐performance antistatic ethylene–vinyl acetate copolymer/high‐density polyethylene composites with graphene nanoplatelets coated by polyaniline.* Journal of Applied Polymer Science, 2017. **134**(37): p. 45303.
33. Wang, D.P. and H.C. Zeng, *Nanocomposites of anatase− polyaniline prepared via self-assembly.* The Journal of Physical Chemistry C, 2009. **113**(19): p. 8097-8106.
34. Abad, B., et al., *Improved power factor of polyaniline nanocomposites with exfoliated graphene nanoplatelets (GNPs).* Journal of Materials Chemistry A, 2013. **1**(35): p. 10450-10457.
35. Geng, Y., S.J. Wang, and J.-K. Kim, *Preparation of graphite nanoplatelets and graphene sheets.* Journal of colloid and interface science, 2009. **336**(2): p. 592-598.
36. David, T., et al., *Part-A: Synthesis of polyaniline and carboxylic acid functionalized SWCNT composites for electromagnetic interference shielding coatings.* Polymer, 2014. **55**(22): p. 5665-5672.
37. Wang, Q., et al., *Preparation of high antistatic HDPE/polyaniline encapsulated graphene nanoplatelet composites by solution blending.* RSC advances, 2017. **7**(5): p. 2796-2803.
38. Mahato, N., N. Parveen, and M.H. Cho, *Synthesis of highly crystalline polyaniline nanoparticles by simple chemical route.* Materials Letters, 2015. **161**: p. 372-374.
39. Yan, J., et al., *Preparation of a graphene nanosheet/polyaniline composite with high specific capacitance.* Carbon, 2010. **48**(2): p. 487-493.
40. Bhadra, S. and D. Khastgir, *Determination of crystal structure of polyaniline and substituted polyanilines through powder X-ray diffraction analysis.* Polymer Testing, 2008. **27**(7): p. 851-857.
41. Sabzi, M., et al., *Graphene nanoplatelets as poly (lactic acid) modifier: linear rheological behavior and electrical conductivity.* Journal of materials chemistry A, 2013. **1**(28): p. 8253-8261.

42. Fitzer, E., et al., *Chemical interactions between the carbon fibre surface and epoxy resins.* Carbon, 1980. **18**(6): p. 389-393.
43. Broyles, N., et al., *Sizing of carbon fibres with aqueous solutions of poly (vinyl pyrollidone).* Polymer, 1998. **39**(12): p. 2607-2613.
44. Wicks, S.S., R.G. de Villoria, and B.L. Wardle, *Interlaminar and intralaminar reinforcement of composite laminates with aligned carbon nanotubes.* Composites Science and Technology, 2010. **70**(1): p. 20-28.
45. Ashori, A., S. Menbari, and R. Bahrami, *Mechanical and thermo-mechanical properties of short carbon fiber reinforced polypropylene composites using exfoliated graphene nanoplatelets coating.* Journal of Industrial and Engineering Chemistry, 2016. **38**: p. 37-42.
46. Chiou, Y.-C., H.-Y. Chou, and M.-Y. Shen, *Effects of adding graphene nanoplatelets and nanocarbon aerogels to epoxy resins and their carbon fiber composites.* Materials & Design, 2019. **178**: p. 107869.
47. Tareq, M.S., et al., *Investigation of the flexural and thermomechanical properties of nanoclay/graphene reinforced carbon fiber epoxy composites.* Journal of Materials Research, 2019. **34**(21): p. 3678-3687.
48. Jang, J., J. Bae, and K. Lee, *Synthesis and characterization of polyaniline nanorods as curing agent and nanofiller for epoxy matrix composite.* Polymer, 2005. **46**(11): p. 3677-3684.
49. Perrin, F.X. and C. Oueiny, *Polyaniline thermoset blends and composites.* Reactive and Functional Polymers, 2017. **114**: p. 86-103.
50. Wei, J., et al., *Graphene nanoplatelets in epoxy system: dispersion, reaggregation, and mechanical properties of nanocomposites.* Journal of Nanomaterials, 2015. **2015**.